\documentclass[12]{article}
\usepackage{epsfig}
\newcommand{\be}{\begin{equation}}
\newcommand{\ee}{\end{equation}}
\newcommand{\bea}{\begin{eqnarray}}
\newcommand{\eea}{\end{eqnarray}}
\newcommand{\no}{\nonumber}

\begin{document}

\begin{flushright}
VLBL Study Group-H2B-8\\
AS-IHEP-2002-030 \\
\today
\end{flushright}
\vskip 3ex

\begin{center}
{\Large Modeling realistic Earth matter density for \\
            CP violation in neutrino oscillation}
\vspace{1cm}

Lian-You Shan$^{a,}$\footnote{shanly@ihep.ac.cn}, 
Yi-Fang Wang, Chang-Gen Yang, Xinmin Zhang \\
\vspace{2ex}
Institute of High Energy Physics, Chinese Academy of Sciences, \\
P.O.Box 918, Beijing 100039, China \\
\vskip 1ex
$^a$CCAST (World Laboratory), P.O.Box 8730, Beijing 100080, China \\
\vskip 3ex
Fu-Tian Liu  \\
Institute of Geology and Geophysics, Chinese Academy of Sciences, \\
P.O.Box 9825 Beijing 100029, China  \\
\vskip 3ex
Bing-Lin Young   \\
Department of Physics and Astronomy, \\Iowa State University,
Ames, Iowa 50011, U.S.A.  
\end{center}
\vspace{2cm}
 
\begin{abstract}
We examine the effect of a more realistic Earth matter density model which 
takes into account of the local density variations along the baseline of a possible 
2100 km very long baseline neutrino oscillation experiment.  Its influence to 
the measurement of CP violation is investigated and a comparison with the 
commonly used global density models made.  Significant differences are
found in the comparison of the results of the different density models.  
\vskip 2ex
\noindent PCAC 14.60.Pq , 11.30.Er
\end{abstract}

\newpage
\section{Introduction}

The first results from KamLAND collaboration have recently been announced. 
The data demonstrate the disappearance of the reactor ${\bar {\nu}_e}$ at a 
high level of confidence \cite{kamland} and hence corroborate the oscillation 
solution to the solar neutrino problem.  Furthermore, the measurement 
excludes all but the Large Mixing Angle oscillation solutions (LMA) 
\cite{snoneutr, Bahcall}.  In the fit of all solar neutrino data, prior the 
KamLAND result, including the neutral and charge currents, elastic scattering 
of both day and night data, the sign of the $\Delta{m}^2_{21}$ has been 
determined to be positive better than 3$\sigma$ for the solar mixing angle, 
$\theta_{\rm solar} < \pi/4$  
\cite{Barger1, Bahcall1}.  Among all the theoretical and 
phenomenological implications \cite{lma}, the LMA solution establishes a 
favorable condition for a determination of the leptonic CP violation (CPV) phase 
\cite{lepcpv} as a further probe of new physics in long baseline neutrino 
oscillation experiments \cite{lbl}.

Because of the smallness of the mixing angle $\theta_{13}$ \cite{Chooz}, 
the signal of CPV effect will not be large in general. This makes the 
measurement of the CP phase a challenging task.  Hence a detailed estimate 
of the various possible theoretical uncertainties and experimental errors will 
be crucial for the extraction of the CP phase.  In particular, the matter effect 
has to be properly delineated in very long baseline (VLBL) oscillation 
experiments.  The presently available Earth matter densities are modeled 
globally averaged values. They are given as functions that depend on the 
Earth radius only.  However, Earth density is not a spherically symmetric  
function independent of the longitudinal and latitudinal coordinates.  There 
are local density variations and can have abrupt density changes from place 
to place, radially or at different longitude and latitude.   
Although the average density model may be suitable for a variety of purposes, 
one wonders if it is adequate for an accurate extraction of the lepton CP phase 
by VLBL experiments.  We can identify two specific questions which may 
affect the outcome of VLBL experiments and to which we have to look for 
answers:  One question is how do we assign a realistic error to the model 
matter density?  The other question is how do we estimate the effect of 
local matter density deviations from the available average?   Unless we find 
satisfactory answers to these questions, we can not be sure that the errors 
in the extraction of the CP phase from VLBL is under control so that we can
assign a good confidence level to the value of the CP phase obtained.

We have addressed the first question in a recent publication~\cite{cpden}
where we proposed a set of density profiles which are randomly distributed 
around the average density to simulate the way that Earth matter density is 
determined.  This approach provides a way to estimate the error, induced 
by the uncertainty of Earth matter density, on the CP phase determination.  
Other approaches have subsequently been proposed and a summary of 
several approaches can be found in~\cite{morecpden}.  

\begin{figure}[t]
\vspace{0.5cm}
\begin{center}
\epsfig{file=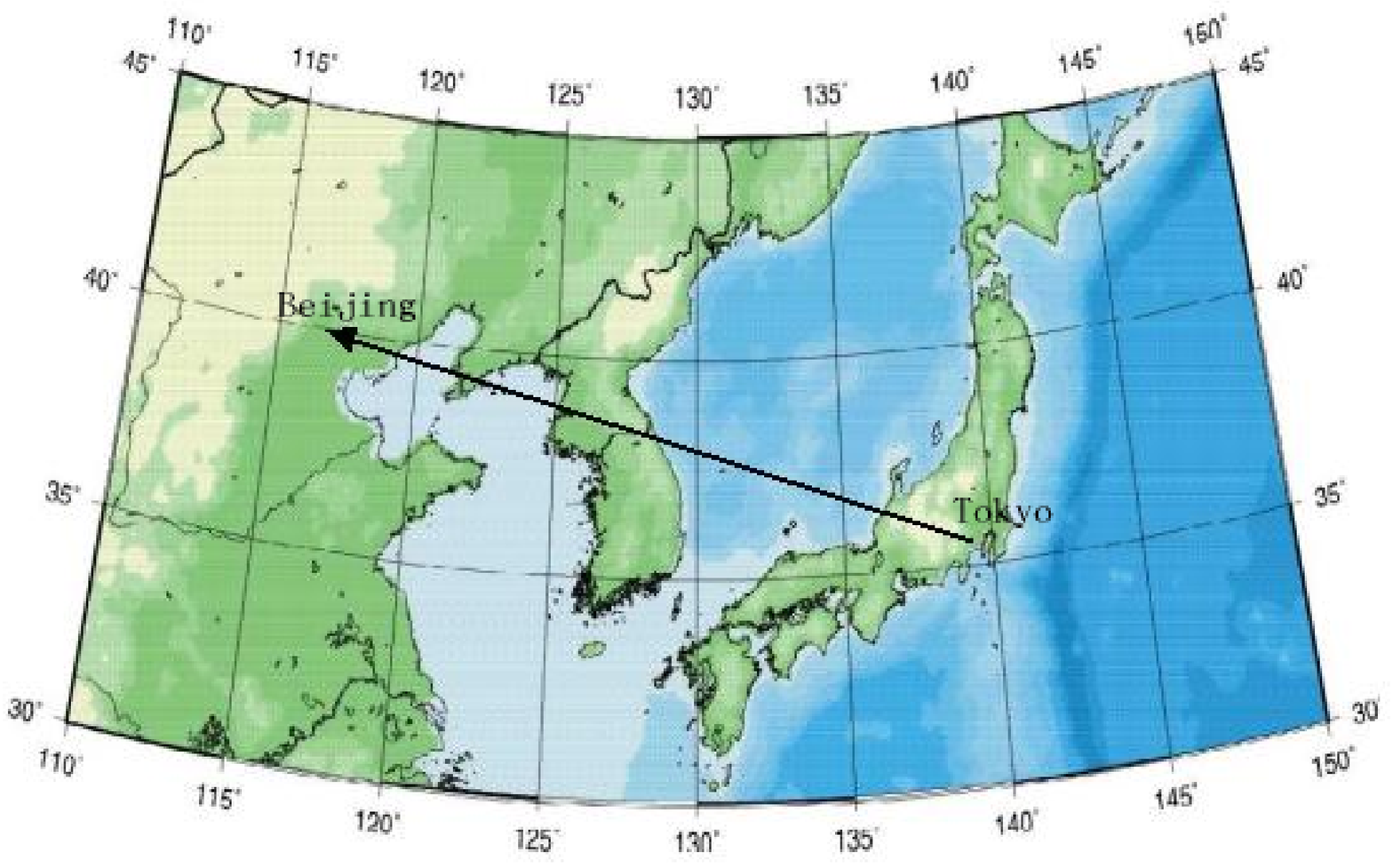,width=16.5cm}
\caption{ 
A schematic diagram of the H2B baseline which consists of a neutrino 
superbeam from J-PARC which is located about 60 km northeast of Tokyo, 
Japan to a detector near Beijing, China. The longitude and latitude of the 
neutrino source and target are indicated.   The baseline is 2100 km.} 
\label{fig:h2beam}
\end{center}
\end{figure}

In this paper we address the second question.  We consider a more realistic 
matter density function along a specific baseline so that we can use a 
concrete example to examine the question.   We will again focus on the recently 
approved high intensity proton synchrotron facility near Tokyo, Japan, i.e., 
the J-PARC (Japan Proton Accelerator Research Complex)~\cite{J-PARC}.  
We consider a VLBL,  with $\nu_\mu$ and $\bar{\nu}_\mu$ beams from
J-PARC to a detector located near Beijing, China, as depicted in  
Fig. \ref{fig:h2beam}.  A preliminary study of the possibility of such a 
VLBL experiment, which we called H2B, can be found in 
Refs.~\cite{h2b} and~\cite{Japan}.  In~\cite{h2b, Japan} and subsequent 
studies a number of physics issues have been investigated: physics potentials
of H2B, relevant backgrounds and errors~\cite{fim}, the effects of Earth 
matter density uncertainties~\cite{cpden}, and the feasibility of measuring 
CP violation and atmospheric neutrino mass ordering in two joint LBL 
experiments \cite{jointcp}.  
Other studies on the matter effect on CP can be found in \cite{mattercp}.
 
In Section 2 we discuss briefly the Earth density models and propose an
alternative, more realistic density model for H2B.  Section III presents a 
quick review of the approach of Ref. \cite{cpden} for dealing with 
uncertainties of Earth matter density.  Section IV discusses briefly the 
general statistical and other errors in CP measurements in LBL.  A brief  
discussion is presented in Sec. V.

\section{Earth density models and matter density distributions}


In looking for the Earth matter effect, we are usually provided with some 
global model of Earth matter density, e.g., the  PREM \cite{prem} or 
AK135 \cite{ak135}. 
All presently available Earth 
density functions are not directly measured but obtained using a limited 
set of geophysics data which are analyzed by means of an inversion 
procedure.  The density so obtained is a function of the depth from the Earth
surface and any longitudinal and latitudinal variations are ignored. Consequently, 
in a given density model, the same density profile will be given for all baselines 
of the same length irrespective of their locality.  All oscillation experiments
of the same baselines length will have the same matter effect. 
Furthermore. since the matter density is a symmetric function along the baseline
the matter effect will be the same when the neutrino beam source and detector
sites are interchanged.  Because of these simplified features, the existing 
density models are inherently of limited level of precision for VLBL.
To improve the precision we have to know the specific density profile for a 
given VLBL.  Or we have to establish the fact that the effect of the density 
variation is within the tolerance of the uncertainties that exist for the experiment.  

\begin{figure}[t]
\vspace{-1.5cm}
\begin{center}
\epsfig{file=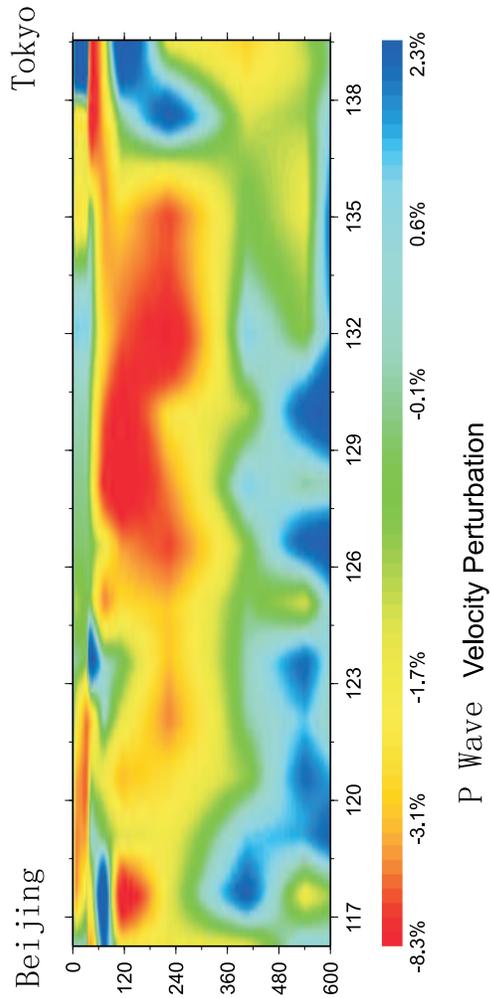, width=9cm }
\vskip 3ex
\caption{ 
A 2-D plot of the earthquake P-wave data based on which the H2B-FTL 
density model is constructed.
The scale from Tokyo to Beijing is the longitude in degrees and the scale 
perpendicular to it is the depth measured from Earth surface in units of km.  
The neutrino trajectory of H2B is a symmetric arc from Tokyo to Beijing, 
with its deepest penetration of Earth about 90 Km near 128$^\circ$ 
longitude. }
\label{fig:denpicture}
\end{center}
\end{figure}

\begin{figure}[t]
\vspace{-1.8cm}
\begin{center}
\epsfig{file=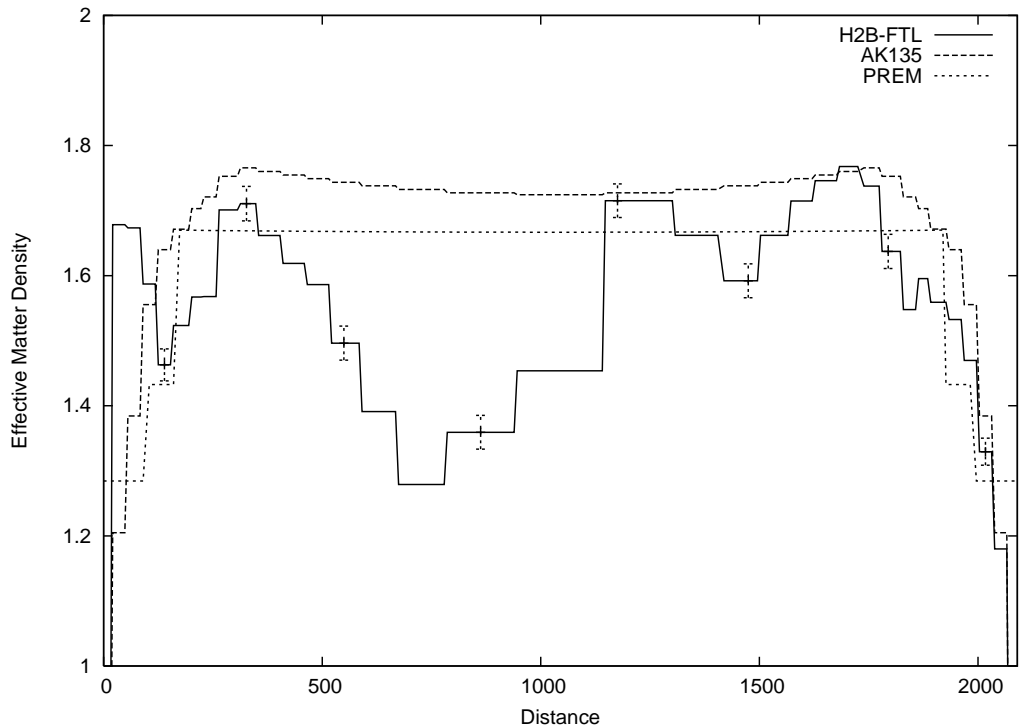,width=14cm}
\caption
{The vertical axis is the Earth effective matter density profile, where
$Y_{\rm e}\rho(x)$, $Y_{\rm e}=0.494$ is the electron fraction and 
$\rho(x)$ is the matter density in gram/cm$^3$.  The horizontal axis 
is the baseline from J-PARC (0 km) to Beijing (2100 km) in km.  The 
PREM and AK135 are world average densities and H2B-FTL is constructed 
specifically for H2B. The error bars on the solid curve are the uncertainties
of H2B-FTL defined in Eq.~(\ref{error}).  Note that the densities have been 
scaled by a factor of the nuclear composition, $Y_{\rm e}=Z/A\sim 0.494$, 
which is suitable in the region of H2B. 
}
\label{fig:denprofile}
\end{center}
\end{figure}

To demonstrate the effect of local density variations we make a detailed 
examination of the mass density profile along the baseline of H2B.  
Figure~\ref{fig:denpicture} shows the Earthquake P-wave velocity 
perturbation around the AK135. The color codes the size of the 
deviations from AK135.  By mapping out the deviations along the baseline,
we can obtain a more realistic density profile for H2B. We shall call this 
density profile for H2B the H2B-FTL density function.\footnote{The 
H2B-FTL density file is based on the work of one of the authors and his 
geophysics research group \cite{FTL863}.}  
We will describe below how H2B-FTL is obtained.  Three density profiles 
along the H2B baseline are shown in Fig. \ref{fig:denprofile}:  PREM by the 
dotted curve, AK135 by the dashed curve, and H2B-FTL together with its 
typical error bars by the solid curve.  The three density profiles provide a 
significant range of density variations which allow us to investigate the effect 
of Earth density on the determination of neutrino oscillation parameters in 
VLBL, in particular the CP phase.  

Similar to PREM and AK135, H2B-FTL is based on the traveling time of 
earthquake waves.  However, it is concentrated on the local region of the 
H2B baseline and therefore involves a much larger set of available data 
along the baseline than either PREM or AL135.  Moreover, since it is a 
3-dimensional density model, it contains significant information in the 
longitudinal and latitudinal directions, while PREM and AK135 are 
1-dimension density models which provide information only along
the radial direction.  Hence H2B-FTL can better represent the actual density 
along the H2B path than either PREM or AK135.

The density function of H2B-FTL is related to that of AK135 via the following
relation \cite{FTL863,density},
\be
\rho_{\rm H2B-FTL} = 
             ( 1 + \frac{1}{k} \frac{\delta v }{v}) \rho_{\rm AK135},
\label{denFTL863}
\ee
where the AK135 density function, $\rho_{\rm AK135}$, can be found 
in~\cite{ak135}.  $v$ is the P-wave velocity and $\delta v$ is the P-wave
velocity correction to AK135~\cite{seismic}.  
The geophysics consideration of the H2B path gives $k=0.2$.  The ratio of 
the P-wave velocity correction to the P-wave velocity is the sole 
geophysics input in the corrections to AK135 and is given in terms of a large 
set of discrete data on the various positions along the baseline~\cite{FTL863}.  
As shown in the scale at the bottom of 
Fig. \ref{fig:denpicture}, $\delta{v}/v$ varies from +2.3\% to -8.3\%.
 
As shown in  
Fig.~\ref{fig:denprofile}, in several regions, H2B-FTL deviates from AK135 
(PREM) beyond the usually cited allowed variation of $\pm 2\%$ ($\pm 5\%$).  
The strongest deviations can be recognized as the red and blue regions of 
Fig. \ref{fig:denpicture}.   The red  and blue regions represent respectively
negative and positive corrections to AK135.  As Fig. \ref{fig:h2beam} displays, 
in its path from J-PARC to Beijing, a neutrino will go through the upper 
Earth mantle which exhibits plastic properties.  It will first experience the 
Japan island crust (blue), the asthenosphere raised by Pacific slab in 
collision with the European-Asian slab (red), the normal asthenosphere 
under the Sea of Japan (red), the bottom of the West-CoSon-Man of west 
Korea peninsula (red), and the Bo Hai Sea of China (red).  As shown in 
Fig.~\ref{fig:denprofile}, global density models, such as PREM and AK135, 
can have significant deviations from the actual mass density profile of 
the H2B baseline. 

The actual variance of the H2B-FTL is a complicate function along the baseline. 
For simplicity and to be conservative, we will ignore its position dependence and 
just take the maximal square root variance \cite{FTL863} to represent the 
density variance,  
\be
\sigma ( \frac{\delta v }{v})= 0.003
\ee
Then we have 
\bea
\frac{ \sigma (\rho_{\rm H2B-FTL} ) }{ \rho_{\rm H2B-FTL} }
   &=& \frac{\rho_{\rm AK135}}{\rho_{\rm H2B-FTL} } \frac{1}{k} 
   \sigma ( \frac{\delta v}{v}) \nonumber \\
   &=& 0.015\times \frac{\rho_{\rm AK135} }{\rho_{\rm H2B-FTL}  }
\label{error}
\eea
which lies between $0.75\%$ and $3\%$.   It should be mentioned that the  
H2B-FTL density profile consists of a huge data sample, hence the 
discretization size of the baseline in the path integral can be as small as 
40 km, while the discretization size of PREM and AK135 along this baseline 
is 200 km.  A smaller discretization size can reduce the error which is 
another factor that contributes to the higher level of precision of H2B-FTL. 

\section{Errors in CP violation measurements}

We review briefly the method of \cite{cpden} and define our notation.  
We are interested in quantifying the possible error caused by Earth density
uncertainties. As usual, we define
a CP-odd difference of neutrino oscillation probability functions,
\bea
D(\delta_{\rm CP}, N_e(x))  &\equiv& 
P_{\alpha\beta}(\delta_{\rm CP}, N_e(x)) - 
P_{{\bar\alpha}{\bar\beta}}(\delta_{\rm CP}, N_{\rm e}(x) )
\label{cpd},
\eea 
where $\delta_{CP}$ is the CP phase, $N_{\rm e}(x)$ the electron 
density distribution function along the baseline $x$, and $P_{\alpha\beta}$
($P_{\bar\alpha\bar\beta}) $ the oscillation probability of 
$\nu_\alpha\rightarrow \nu_\beta$ ($\bar\nu_\alpha\rightarrow \bar\nu_\beta$).
The electron density function  $N_{\rm e}(x)$ is related to the 
Earth matter density $\rho( x )$ by the Avogadro number, $N_{\rm A}$, 
and the electron fraction, $Y_{\rm e}$, through the usual relationship, 
$N_{\rm e}(x) = N_{\rm A} Y_{\rm e}\rho(x)$.  
The dependence on the neutrino energy, mixture angles, and neutrino masses is 
suppressed.  

We estimate the possible error from density uncertainty by following the 
formalism of ~\cite{cpden}, 
\bea
\delta D(\delta_{\rm CP},N_{\rm e}(x))  
      & \equiv & 
      \sqrt{ < { \left( (P_{\alpha\beta} -  P_{{\bar\alpha}{\bar\beta}}) 
   - < (P_{\alpha\beta}- P_{{\bar\alpha}{\bar\beta}}) > \right) }^2 >} \no \\
& = & \sqrt{ \left(\delta (P_{\alpha\beta})\right)^2              
                 +  \left(\delta (P_{{\bar\alpha}{\bar\beta}})\right)^2},
\label{VarCPN}
\eea
where $<\cdot\cdot\cdot>$ denotes a weighted average of a matter density 
dependent quantity.  It is defined as a path integral of the quantity, along a 
given baseline, over an ensemble of possible variations of Earth matter 
density profiles, $N_{\rm e}(x)$, weighted by a logarithmic normal 
distribution functional $F[N_e(x)]$ for non-negative quantities 
\cite{cpden}, e.g.,   
\bea
< P_{\alpha\beta} > & \equiv &
\int [{\cal D} N_{\rm e}(x)]  F[N_{\rm e}(x)] 
                    P_{\alpha\beta}(\delta_{CP}, N_{\rm e}(x)).
\label{PP}
\eea
Therefore we interpret the average matter density $\hat{N}_{\rm e}(x)$, such 
as PREM and AK135, as a weighted average over all samples of possible density 
profiles,
\bea
\hat{N}_{\rm e}(x) &\equiv& <N_{\rm e}(x)>
       = \int [{\cal D}N_{\rm e}(x)] F[N_{\rm e}(x)] N_{\rm e}(x). 
\eea
The matter density uncertainty is given as usual by
\bea 
\sigma (x) &\equiv& \sqrt{<N^2_{\rm e}(x)> -
                                        \left(\hat{N}_{\rm e}(x)\right)^2}
\eea
The level of precision of a density model with a given $\hat{N}(x)$ and
$\sigma(x)$ can be measured by 
\bea
r(x)  & \equiv & \frac{\sigma(x)}{{\hat N}_{\rm e}(x)}.
\eea

To do the functional integral, the integration path along the neutrino 
baseline is discretized according to the available geophysical information. 
For the H2B baseline, the discretization size for H2B-FTL is 40 km, 
which is much smaller than the 200 km discretization size suitable for 
PREM or AK135.  The smaller discretization size is helpful in reducing 
the errors in the functional integration.

We remark that an alternative parameterization of the Earth matter density
uncertainty is to give the average matter density a fixed deviation, i.e., taking
the density function to be $N'(x) = (1\pm r'(x))\hat{N}(x)$, where $r'(x)$ is 
the density uncertainty. Conventionally, $\hat{N}(x)$ is given by PREM or 
AK135 and $r'(x)$ is respectively 0.05 or 0.02. As discussed in~\cite{cpden}, 
this parameterization will lead to a larger uncertainty in the extraction of the 
CP phase than the present approach.


The above formulation can be readily adopted to the measurable quantity
of the event number.  Unless noted otherwise in the calculation of event
number, we assume 6500 interacting muon neutrinos for H2B. 

\begin{figure}
\begin{center}
\epsfig{file=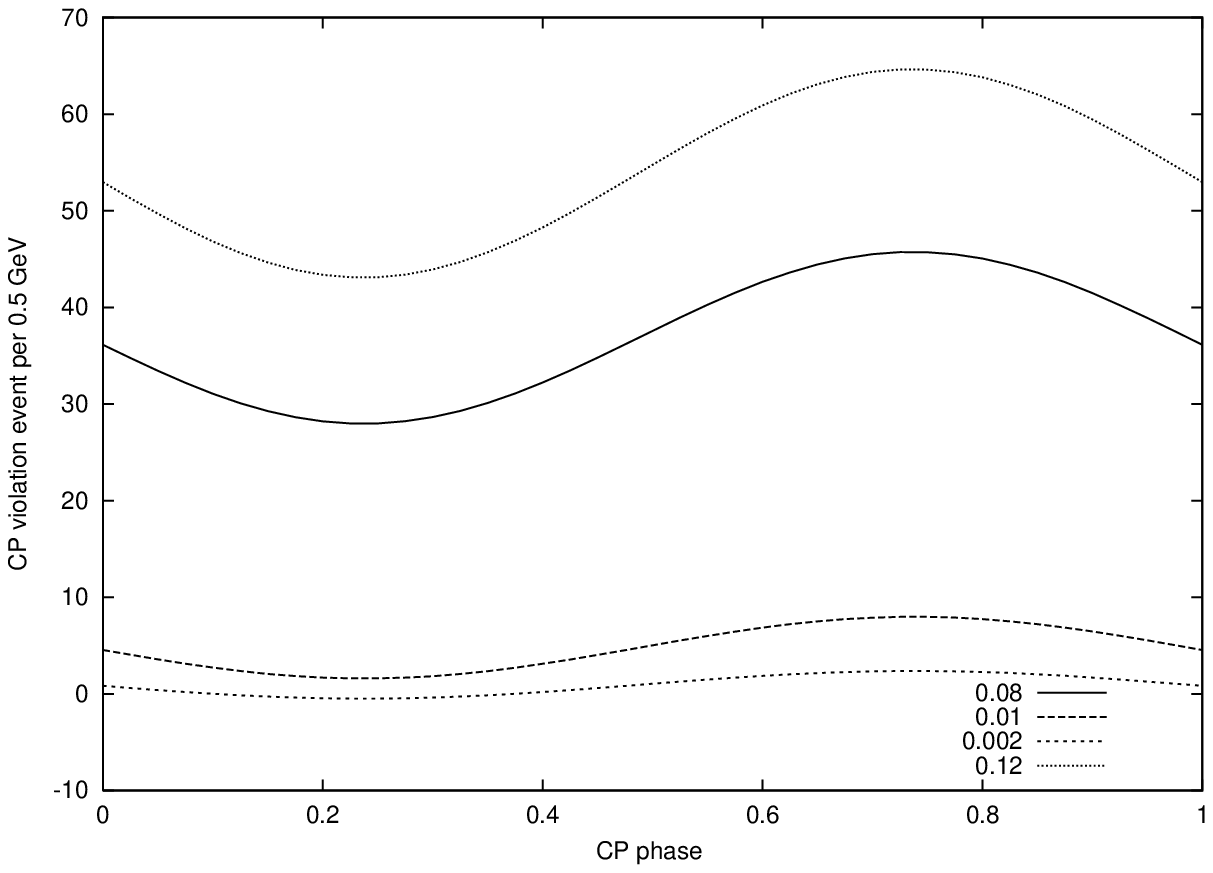,width=15cm}
\epsfig{file=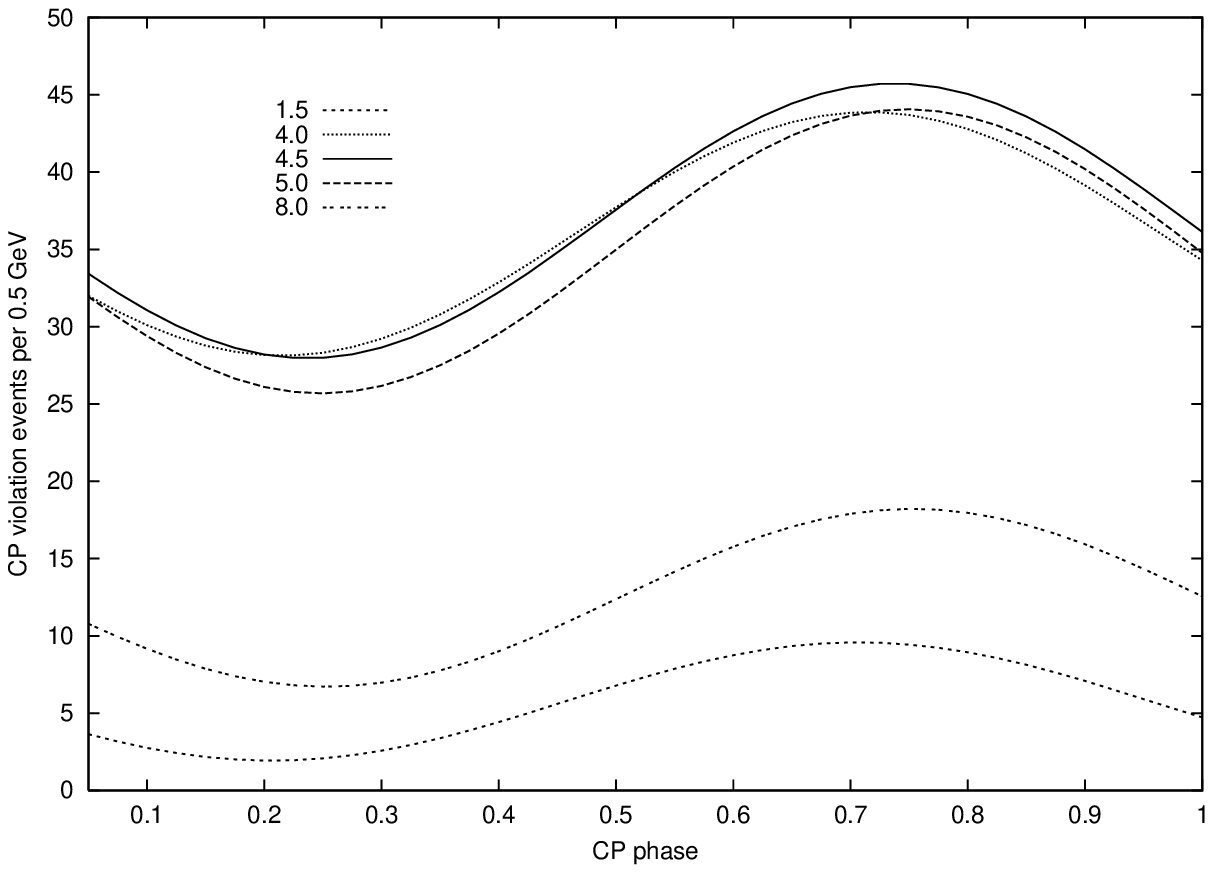, width=15cm }
\caption{
The $\nu_{\rm e}$ and $\bar{\nu}_{\rm e}$ number difference, 
corresponding to Eq.~(\ref{cpd}) with 6500 interacting $\nu_\mu$,  
vs the leptonic CP phase. The curves in the upper panel are for different 
values of $\sin^2 2\theta_{13}$ and those in the lower panel are for different 
energies. The horizontal axis which is the CP phase $\delta_{\rm CP}$ is
in units of $2\pi$}
\label{fig:cpvenergy}
\end{center}
\end{figure}

For the numerical calculation, we adopted the following values for the 
solar and atmospherical neutrino mass square differences and the corresponding
mixing angles:\footnote{The recent KamLAND best fit \cite{kamland} gives
$\Delta{m}_{\rm atm}^2 =6.9\times 10^{-5}$ and 
$0.86\leq \sin^22\theta_{12}\leq 1.0$.  The value of $\sin^22\theta_{12}$ 
used in the present work corresponds to the lower limit of the KamLAND 
value.}
\bea
\Delta m^2_{\rm solar} &=& 5.0\times {10}^{-5}~{\rm eV^2}, 
         \hskip 2ex \tan^2 \theta_{12} ~=~ 0.42 ,  \no  \\
\Delta m^2_{\rm atm} &=& 3.0 \times {10}^{-3}~{\rm eV^2}, 
        \hskip 2ex \sin^2 2\theta_{23} ~=~ 0.99. 
\eea
Since the CPV effect is proportional to $\theta_{13}$, a larger value of 
$\sin^2 2\theta_{13}$ will be more favorable for its measurement.  The 
CHOOZ bound is $\sin^22\theta_{13}\leq 0.15$~\cite{Chooz}.  In 
Fig.~\ref{fig:cpvenergy} we plot the CPV event number vs the CP phase.
We show several different values of $\sin^22\theta_{13}$ and the beam
energies. One can see from the upper panel Fig. \ref{fig:cpvenergy} that, 
for this number 
of testing muons there will be little sensitivity in the CP phase measurement in 
H2B  if $\sin^2 2\theta_{13}<0.01$.  However with a higher number of testing 
muons the sensitivity can be increased.  In the following calculation we will use 
$\sin^2 2\theta_{13}=0.08$ for illustration.  However, for comparison, we will 
also show some results for $\sin^2 2\theta_{13}=0.01$.   

To select the appropriate neutrino energy, we have examined the oscillation
probabilities of 1.5, 4, 4.5, 5, and 8 GeV.
 As shown in the  lower panel of Fig. \ref{fig:cpvenergy}, 
$4.5$ GeV is the optimal energy which will be used in 
all subsequent calculations.  It should be noted that 
Fig. \ref{fig:cpvenergy} 
employs the density distribution H2B-FTL, but different density model does 
not change the optimal energy and the sensitivity in $\theta_{13}$.
  
\begin{figure}[t]
\begin{center}
\epsfig{file=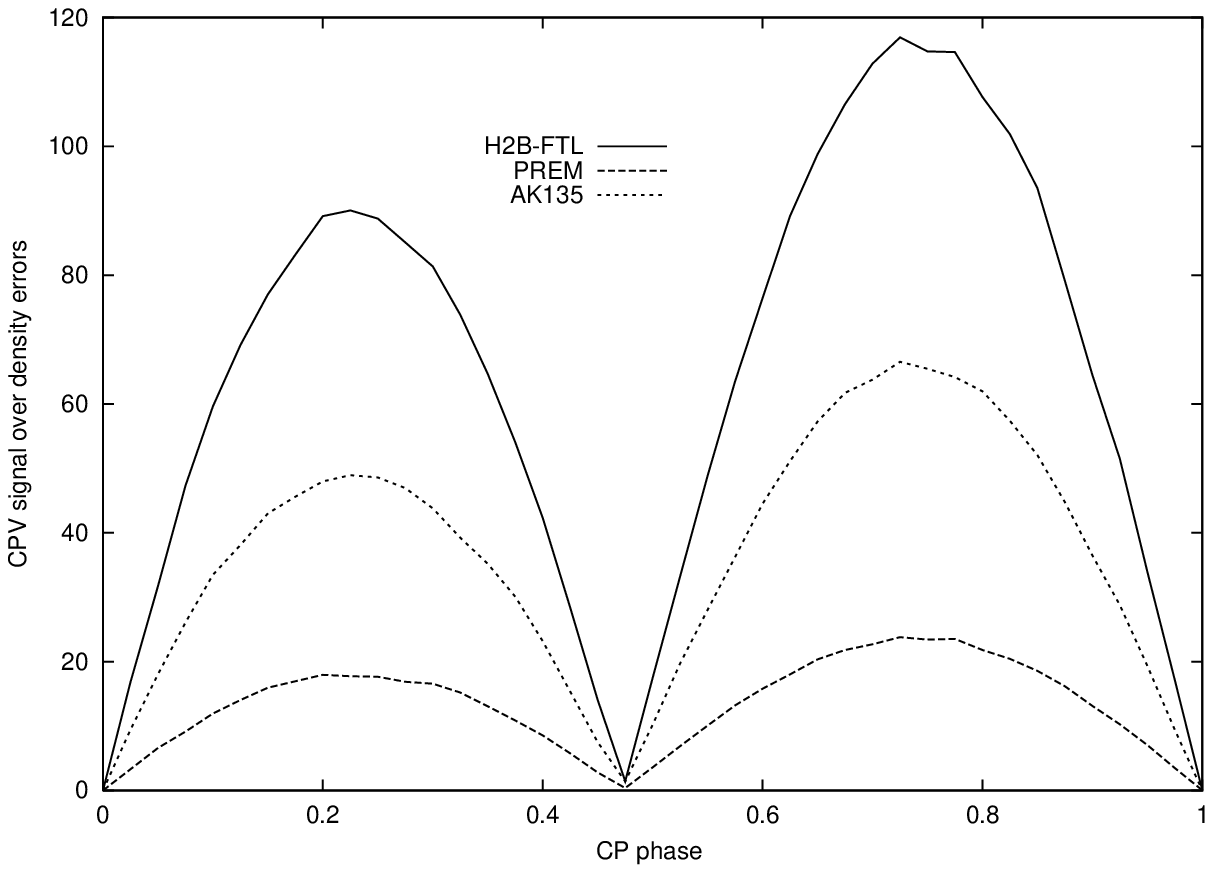,width=14.5cm}
\epsfig{file=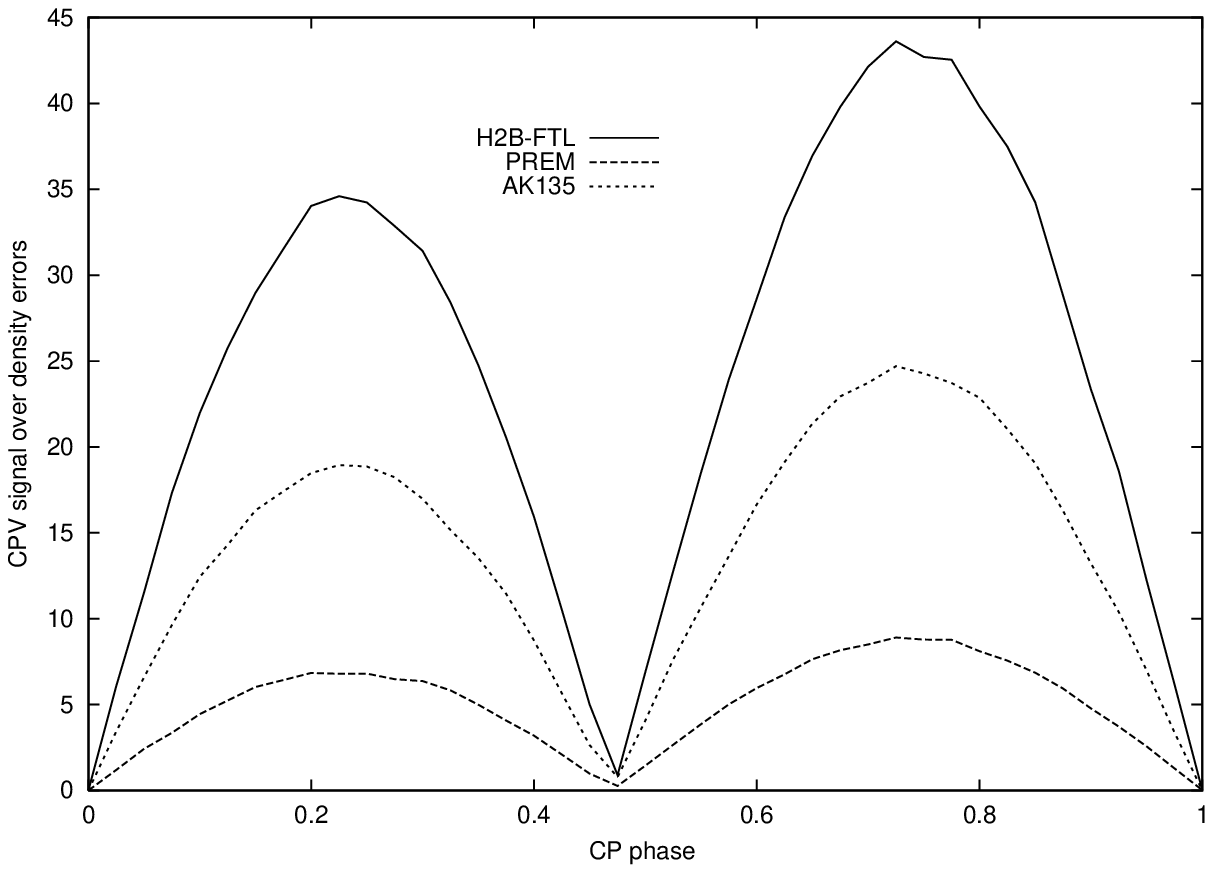,width=14.5cm}
\caption{
The CPV signal significance $S_{\rm den}$, Eq. (\ref{signif}), vs the CP 
phase. The solid line is for H2B-FTL with a $3\%$ uncertainty, the 
short-dashed line is for AK135, and the dashed line for PREM.  The upper 
panel has $\sin^2 2\theta_{13}=0.01$ and the lower panel 
$\sin^2 2\theta_{13}=0.08$.  The scale of the horizontal axis is in units of 
2$\pi$.}
\label{fig:signifvsensimods}
\end{center}
\end{figure}

To examine the difference among the different density models we define the 
following quantity which provides a gross measurement of the "pure" CP effect, 
\be
\Delta{D}(\delta_{\rm CP}) \equiv 
   D(\delta_{CP},N_{\rm e}(x)) - D(\delta_{CP}=0,N_{\rm e}(x)).
\label{sigal}
\ee
Then we can define a quantitative measure of the significance of the CPV signal 
by the ratio of the "pure" CPV effect and the corresponding quantity which also
contains the matter effect\footnote{In Ref. \cite{cpden} (see Eq. (17) there) we 
used the inverse of $S_{\rm den}(\delta_{\rm CP})$ as defined in 
Eq. (\ref{signif}).  Since $\Delta D(\delta_{\rm CP})$ vanishes in the absence 
of CPV, we find the present definition is more convenient to graph.},
\be
S_{\rm den}(\delta_{\rm CP}) \equiv 
     \frac{\Delta D(\delta_{\rm CP})} 
            {\delta D(\delta_{\rm CP},N_{\rm e}(x))}  
\label{signif}
\ee
A larger $S_{\rm den}(\delta_{\rm CP})$ gives a stronger CP signal relative 
to the matter effect, we require that $S_{\rm den}(\delta_{\rm CP})$ be 
significantly greater than 1.  In Fig. \ref{fig:signifvsensimods}, we plot 
$S_{\rm den}(\delta_{\rm CP})$ against the CP phase $\delta_{\rm CP}$.
Indeed H2B-FTL gives a signal better than AK135 and PREM for 
$\delta_{CP}$ away from 0 and $\pi$ where the CPV effect vanishes. 

\begin{figure}[t]
\begin{center}
\epsfig{file=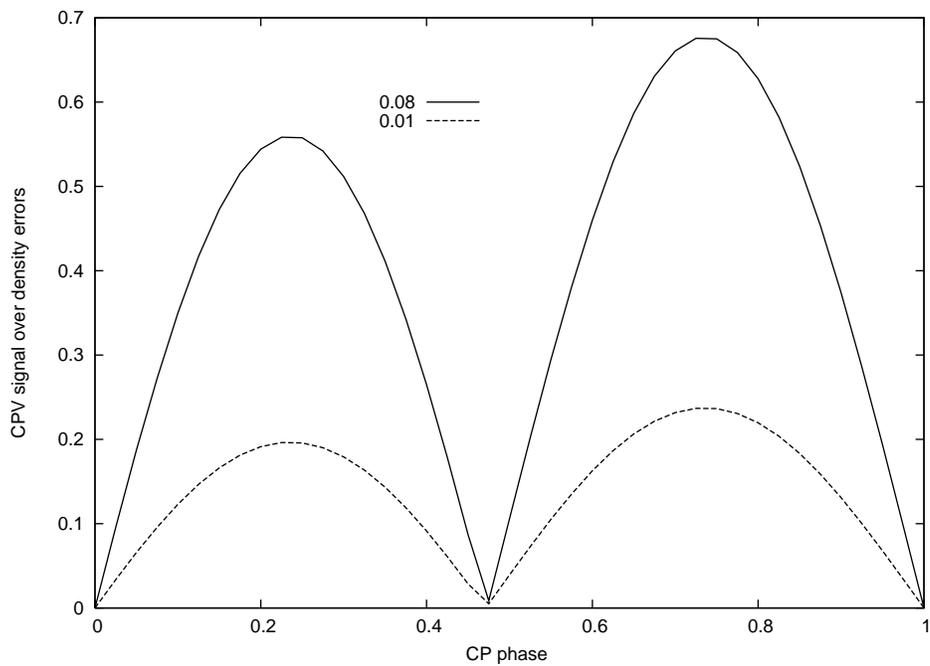}
\caption{
The CPV signal significance $\tilde{S}_{\rm den}$, Eq.~(\ref{truesigal}),
vs the CP phase.  The solid curves is for $\sin^2 2\theta_{13}=0.01 $ and 
dashed curve for $\sin^2 2\theta_{13}=0.08$.  The horizontal axis is in
units of 2$\pi$.  
}
\label{fig:modelmisu}
\end{center}
\end{figure}

To compare the H2B-FTL with other density model, e.g., the AK135,
we define
\bea
\tilde{\delta}D (\delta_{CP})  &=&
D(\delta_{\rm CP},N_{\rm e}(x))_{({\rm H2B-FTL})}
-D(\delta_{\rm CP},N_{\rm e}(x))_{({\rm AK135})}.
\label{trueerror}
\eea
We can define a significance measure of the CPV effect with 
the difference of density models,
\bea
\tilde{S}_{\rm den}(\delta_{\rm CP}) &=& 
         \frac{\Delta D(\delta_{\rm CP})}
                {\tilde{\delta}{D}(\delta_{\rm CP})}.
\label{truesigal}
\eea
We plot $\tilde{S}_{\rm den}(\delta_{\rm C})$ in Fig.~\ref{fig:modelmisu}.
We see that the difference in density models is always larger than the CP 
difference.  This again underlines the fact that a realistic density is necessary. 

\begin{figure}[t]
\vspace{-1.5cm}
\begin{center}
\epsfig{file=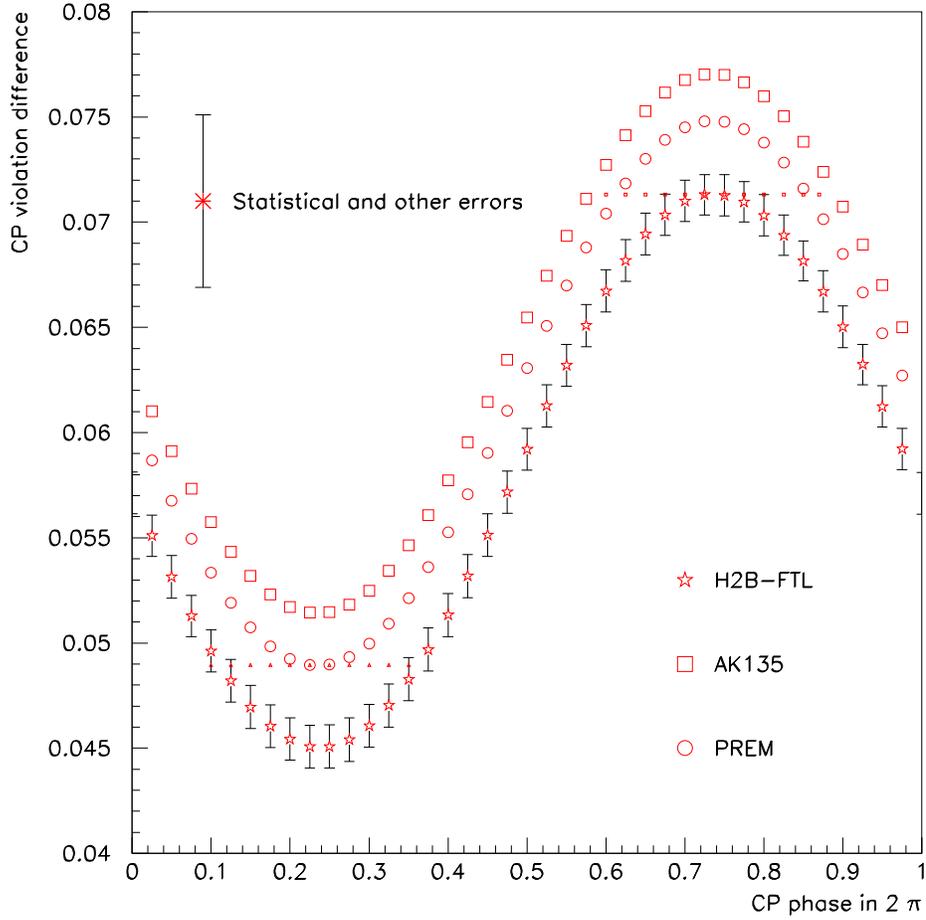,width=14cm}
\caption{
The CPV difference $\delta D(\delta_{\rm CP})$, Eq.~(\ref{VarCPN}), vs 
the CP phase for $\sin^2 2\theta_{13}=0.08$.  The squares are for AK135, 
the circles for PREM, and the stars for H2B-FTL.  The error bars associated
with the stars are for H1B-FTL.  A typical size of the statistical plus background 
errors is shown in the upper left corner.  See the relevant discussion in Sec. 4. 
}
\label{fig:CPVvsdensimods}
\end{center}
\end{figure}

In Fig. \ref{fig:CPVvsdensimods} we plot the "pure" CP effect, 
$\Delta D(\delta_{\rm CP})$ given in Eq. (\ref{sigal}), as a function of the 
CP phase for the three density models together with the error bars for 
H2B-FTL.  In general, AK135 is 3$\sigma$ away and PREM is 6$\sigma$ 
away from H2B-FTL for a given CP phase, indicating significant differences 
between the different density models. 
 
\section{ Statistical and other Errors in CPV measurements}

In this section we present the result of our study of the statistical errors,
the background effects, and the systematic uncertainties, and contrast them 
with the effect of density uncertainties.  
Our treatment of statistical errors, background effects, and systematic
uncertainties follows that of Ref.~\cite{fim}.  They are represented by
their respective error factors denoted as $f$ (statistical), $r$ (background),
and $g$ (systematic).   Denoting the variance of their combined effect by
$\sigma_{\rm SBS}$, we can define the significance 
of measure of the CPV effect with respect to this combined variance,
\bea
S_{\rm SBS} &\equiv& 
    \frac{\Delta D(\delta_{\rm CP })}{\sigma_{\rm SBS}} 
\label{signalSBS}
\eea
Obviously $S_{\rm SBS}$ is only meaningful for the case of a sufficiently large 
number of interacting $\nu_\mu$'s.
For the numerical calculation we take the statistical factor $f=0.02$, the
background factor $r=0.01$, and the systematic factor $g=0$ as used 
in~\cite{jointcp}.
We show in the upper left corner of Fig.~\ref{fig:CPVvsdensimods} 
a representative error bar for this combined error.     

\begin{figure}[t]
\begin{center}
\epsfig{file=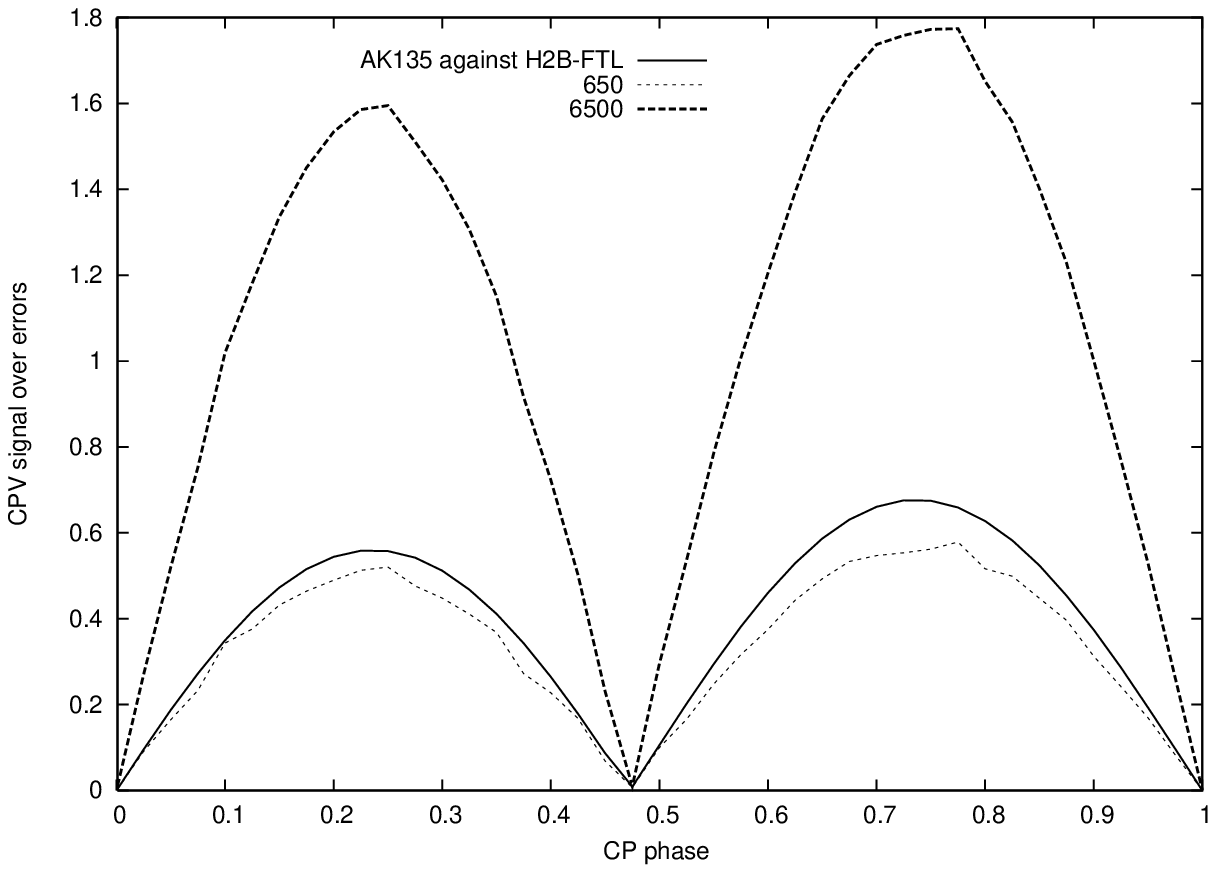,width=14.5cm}
\epsfig{file=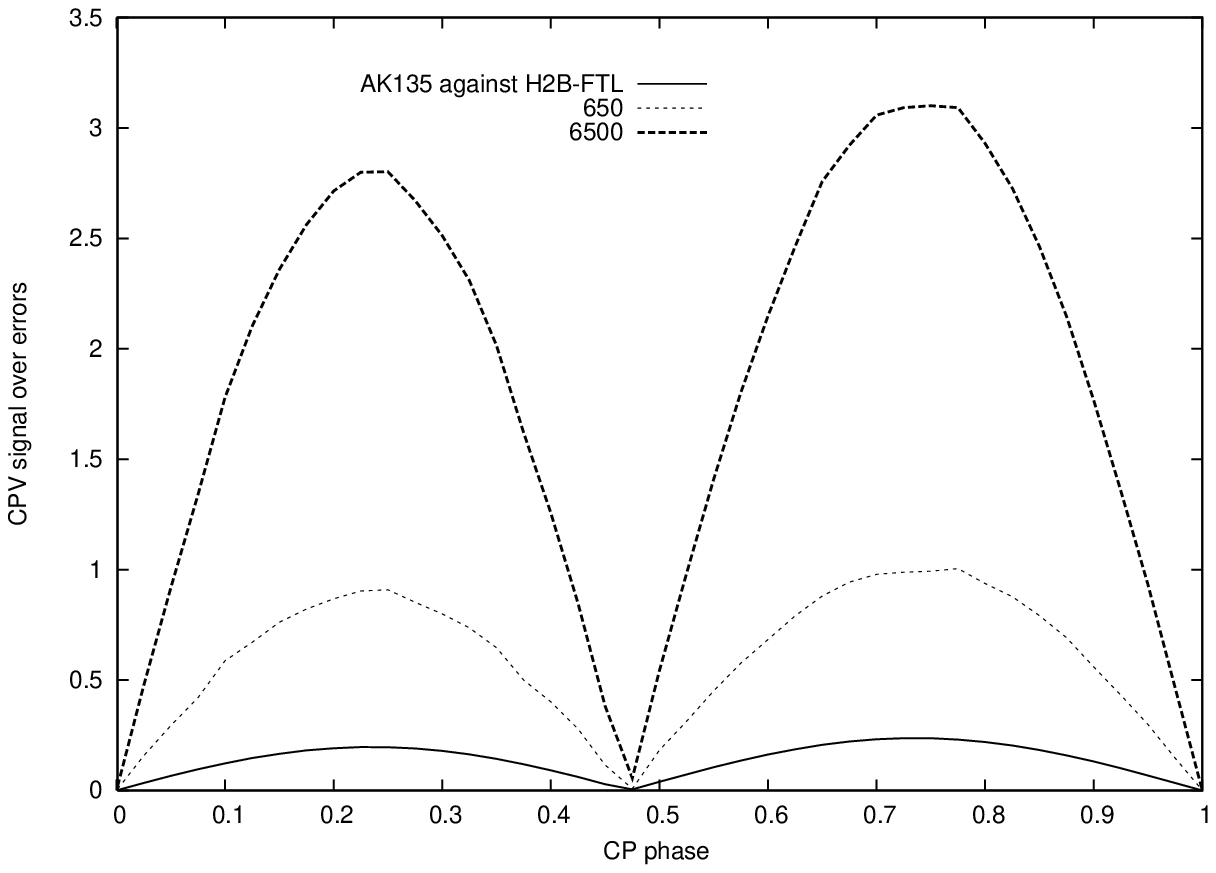,width=14.5cm}
\caption{
The CPV signal significance vs the CP phase.  The solid line is for 
$\tilde{S}_{\rm den}$, Eq.~(\ref{truesigal}).  The dashed lines are 
for the signal significance $S_{\rm SBS}$, Eq.~(\ref{signalSBS}).  The thin 
dashed line has 650 interacting $\nu_\mu$  and the heavy dashed line has
6500 interacting $\nu_\mu$.  The upper panel is for $\sin^2 2\theta_{13}=0.01$
while the lower panel for $\sin^2 2\theta_{13}=0.08$
}
\label{fig:stabackgd}
\end{center}
\end{figure}

In Fig. \ref{fig:stabackgd} we plot both $S_{\rm SBS}$, Eq. (\ref{signalSBS}),
and $\tilde S_{\rm den}$, Eq. (\ref{truesigal}), for 
$\sin^22\theta_{13} = 0.08$ and 0.01 and for two total number of interacting
$\nu_\mu$'s, 6500 and 650.  We see that for several hundred interacting 
$\nu_\mu$'s, no CPV effect is expected to be measurable.  
For several thousands of $\nu_\mu$ muons, we have a chance to observe
the CPV effect.  
With the CPV variable we consider here Earth matter uncertainty can be 
ignored safely if H2B-FTL model is employed, but the significance will 
be worse if other density model ($\tilde S_{\rm den} $) is adopted.


\section{Discussions}
In this paper we investigated the potential error caused by the uncertainties 
of earth matter density.  We also consider the other errors, such as the 
statistical error, in the context of intended H2B physics.
We found that the more realistic density model, H2B-FTL, the matter density
variation induces a rather small error which allows a meaningful separation
of H2B-FTL from AK135 and PREM.   With CPV variable we used, i.e., 
the CP difference, the statistical and other errors generally dominate over the
difference of the density models. Therefore, a different approach of the CP
measurement, such as that used in~\cite{jointcp} has to be investigated.

\newpage
\begin{thebibliography}{9999}
\bibitem{kamland} KamLAND Collab., K. Eguchi et al., Phys. Rev. Lett. 
           {\bf 90}, 021802 (2003) (arXiv: hep-ex/0212021).
\bibitem{snoneutr}SNO Collab., Q.R. Ahmad et al., Phys. Rev. Lett.
            {\bf 87}, 071301 (2001) (arXiv:  nucl-ex/0106015); Phys. Rev. 
            Lett. {\bf 89}, 011302 (2002) (arXiv: nucl-ex/0204009).

\bibitem{Bahcall}
            J.N. Bahcall, M.C. Gonzalez-Garcia, and Carlos Pena-Garay,
            JHEP 0207:054 (2002) (arXiv: hep-ph/0204314)

\bibitem{Barger1} V. Barger, D. Marfatia, K. Whisnant and B.P Wood,
            Phys. Lett. {\bf B573} (2002) 179.

\bibitem{Bahcall1} J.N. Bahcall, M.C. Gonzalez-Garcia, and Carlos
            Pe{\~a}-Garay, {\it Solar Neutrinos before and after KamLAND},
            arXiv: hep-ph/0212147.

\bibitem{lma} C. D. Froggatt, H. B. Nielsen, and Y. Takanishi, 
  Nucl. Phys., {\bf B631} 285-306 (2002) (arXiv: hep-ph/0201152);
  O. L. G. Peres and A. Yu. Smirnov, Nucl. Phys. Proc. Suppl. 
  {\bf 110} 355 (2002) (arXiv: hep-ph/0201069);
  Xiao-Jun Bi and Yuan-Ben Dai, arXiv: hep-ph/0204317;
  S. Antusch, J. Kersten, M. Lindner, and M. Ratz, Phys. Lett. 
  {\bf B544}, 1 (2002) (arXiv: hep-ph/0206078); 
  Zhi-zhong Xing, Phys. Lett. {\bf B550}, 178 (2002) 
 (arXiv: hep-ph/02102776).

\bibitem{lepcpv} T. Endoh, S. Kaneko, S.K. Kang, T. Morozumi, and
   M. Tanimoto,  Phys. Rev. Lett., {\bf 89} 231601 (2002) 
   (arXiv: hep-ph/0209020);
   John Ellis and Martti Raidal, Nucl. Phys. {\bf B643}, 229 (2002)
   (arXiv: hep-ph/0206174);
   Y. Farzan and A. Yu. Smirnov, Phys. Rev., {\bf D65}, 113001 (2002)
    (arXiv: hep-ph/0201195);
    M. C. Gonzalez-Garcia, Y. Grossman, A. Gusso, and Y. Nir, 
    Phys. Rev. {\bf D64} 096006 (2001) (arXiv: hep-ph/0105159);
    J. Burguet-Castell, M.B. Gavela, J.J. Gomez-Cadenas, 
    P. Hernandez, and O. Mena, Nucl. Phys. {\bf B608}, 301 (2001)
    (arXiv: hep-ph/0103258); 
    S. Pastor, J. Segura, V.B. Semikoz, and J.W.F. Valle, Nucl. Phys. 
    {\bf B566}, 92 (2000) (arXiv: hep-ph/9905405). 
    K. Dick, M. Freund, M. Lindner, and A. Romanino, Nucl. Phys.,
    {\bf B562}, 29 (1999) (arXiv: hep-ph/9903308).

\bibitem{lbl} BNL Neutrino Working Group:hep-ex/0211001;
    J. Burguet-Castell, M.B. Gavela, J.J. Gomez-Cadenas,
    P. Hernandez, and O. Mena, Nucl. Phys., {\bf B646}, 301 (2002) 
    (arXiv: hep-ph/0207080);
    Y. F. Wang, K. Whisnant, Zhaohua Xiong, Jin Min Yang, and 
    Bing-Lin Young, Phys. Rev., {\bf D65} (2002) (2002) 
    (arXiv: 0111317);
     K. Dick, Martin Freund, Patrick Huber, and Manfred Lindner, 
     Nucl. Phys. {\bf B598}, 543 (2001) (arXiv: hep-ph/0008016); 
     V. Barger, S. Geer, R. Raja, and K. Whisnant, Phys. Rev. 
     {\bf D62} 013004 (2000) (arXiv: hep-ph/9911524) 

\bibitem{Chooz} CHOOZ Collaboration, M. Apllonio et al., Phys. Lett., 
      {\bf B466}, 415 (1999) (arXiv: hep-ex/9907037). 

\bibitem{cpden}Lian-You Shan, Bing-Lin Yong, and Xinmin Zhang,
    Phys. Rev., {\bf D66} 053012 (2002) (arXiv: hep-ph/0110414). 

\bibitem{morecpden}B. Jacobsson, T. Ohlsson, H. Snellman, and
    W. Winter, Phys. Lett., {\bf B532}, 259 (2002) 
    (arXiv: hep-ph/0112138);
     B. Jacobsson, T. Ohlsson, H. Snellman, and W. Winter, talk given
     at NuFact '02, London, arXiv: hep-ph/0209147.

\bibitem{J-PARC} For a description of J-PARC, we refer to its website
     http://jkj.tokai.jaeri.go.jp/.  Previously the facility has been referred
     to by the acronym JHF or HIPA.

\bibitem{h2b}
      H.S. Chen, et al., {\it Report of a Study on H2B, Prospect of a very 
      long baseline neutrino oscillation experiment HIPA to Beijing}, 
      VLBL Study Group-H2B-1, IHEP-EP-2001-01, May 29, 2001,
      arXiv: hep-hp/0104266

\bibitem{Japan}  M. Aoki, K. Hagiwara, Y. Hayato, T. Kobayashi, 
           T. Nakaya, K. Nishikawa (Kyoto U.), N. Okamura, 
           "{\it Prospects of very long baseline neutrino oscillation
              experiments with the KEK/JAERI high intensity proton 
              accelerator}", KEK-TH-798, VPI-IPPAP-01-03, Dec 2001, 
              arXiv: hep-ph/0112338; and talk given at the 4th International 
              Conference on B Physics and CP Violation (BCP 4), Ago 
              Town, Mie Prefecture, Japan, 19-23 Feb 2001,
               arXiv: hep-ph/0104220. 

\bibitem{fim}Yi-Fang Wang, K. Whisnant, and Bing-Lin Young,
    Phys. Rev., {\bf D65} 073006 (2002) (arXiv: hep-ph/0109053).

\bibitem{jointcp}K. Whisnant, Jin Min Yang, and Bing-Lin Young, 
            Phys. Rev. {\bf D67} (2003) 013004  
            (arXiv: hep-ph/0208193).

\bibitem{mattercp}
    H. Yokomakura, K. Kimura, and A. Takamura, 
    Phys. Lett. {\bf B544}, 286 (2002) (arXiv: hep-ph/0207174); 
    T. Hattori, Tsutom Hasuike, and Seiichi Wakaizumi, Phys. Rev.,
    {\bf D65} 073027 (2002) (hep-ph/0109124);
    T. Ohlsson and H. Snellman, Eur. Phys. J. {\bf C20}, 507 (2001)
     (arXiv: hep-ph/0103252). 

\bibitem{prem}A.M. Dziewonsky and D.L. Andson, Phys. Earth Planet 
    Int., {\bf 25}, 507 (1981).  

\bibitem{ak135}
   B.L.N. Kennet et al., Geophys. J. Int., {\bf 122}, 108 (1995); 
   J.P. Montagner et. al., Geophys. J. Int., {\bf 125}, 229 (1995).

\bibitem{FTL863}
   Fu-Tian Liu et al.,
   {\it Three-dimensional seismic tomography of oceanic lithosphere},
     Project 863 Study report, ( 863-820-01-04 );
     Fu-Tian Liu et al., Geophys. J. Int., {\bf 101}, 379 (1990). 

\bibitem{density}
    K.E. Bullen, {\it The Earth's density}, Chapman and Hall , London (1975).

\bibitem{seismic}F.T. Liu and A. Jin, {\it Seismic Tomography of China},
    in Seismic Tomography: Theory and Practice, Ed. by 
     H.M. Iyer and K. Hiahara, Chapman and Hall , London (1993).
\end{thebibliography}
\end{document}